\documentclass[a4paper]{aa}

\usepackage{natbib}
\usepackage{txfonts}
\usepackage{epsfig}

\begin{document} 

\newcommand{\es}{erg s$^{-1}$}   
\newcommand{\ecms}{erg~cm$^{-2}$~s$^{-1}$}
\newcommand{\halpha}{H$\alpha$}  
\newcommand{\hbeta}{H$\beta$}
\newcommand{\kms}{km~s$^{-1}$}   
\newcommand{\cmthree}{cm$^{-3}$}
\newcommand{\msun}{M$_{\odot}$} 
\newcommand{\xmm}{XMM-\emph{Newton}} 
\newcommand{\nh}{\mbox{$N({\rm H})$}}
\newcommand{\chandra}{\emph{Chandra}}

\title{\chandra\ observations of the massive star-forming region
  S106}\subtitle{X-ray emission from the embedded massive protostellar
  object IRS~4}

\author{G.\ Giardino\inst{1} \and F.\ Favata\inst{1} \and G.\ 
  Micela\inst{2}}

\institute{Astrophysics Division -- Research and Science Support
  Department of ESA, ESTEC, 
  Postbus 299, NL-2200 AG Noordwijk, The Netherlands
\and
  INAF --Osservatorio Astronomico di Palermo, 
  Piazza del Parlamento 1, I-90134 Palermo, Italy 
}

\offprints{G. Giardino (Giovanna.Giardino@rssd.esa.int)}

\date{Received date / Accepted date}

\titlerunning{}
\authorrunning{}

\abstract{We present \chandra\ observations of the massive
   star-forming region S106, a prominent H\,{\sc ii} region in Cygnus,
   associated with an extended molecular cloud and a young cluster.
   The nebula is excited by a single young massive star located at the
   center of the molecular cloud and the embedded cluster.  The
   prominence of the cluster in the \chandra\ observation presented
   here confirms its youth and allows some of its members to be
   studied in more detail.  We detect X-ray emission from the young
   massive central source S106~IRS~4, the deeply embedded central
   object which drives the bipolar nebula with a mass loss rate
   approximately 1--2 orders of magnitude higher than main sequence
   stars of comparable luminosity. Still, on the basis of its wind
   momentum flux the X-ray luminosity of S106~IRS~4 is comparable to
   the values observed in more evolved (main sequence and giant)
   massive stars, suggesting that the same process which is
   responsible for the observed X-ray emission from older massive
   stars is already at work at these early stages.}

\maketitle

\section{Introduction}
\label{sec:intro}

The study of regions of recent star formation provides the
observational data necessary to address a number of important
astrophysical questions, such as the physics of the star formation
process, what determines its efficiency, the characterization of the
initial mass function, the stellar chemical evolution and the galactic
chemical recycling process; in particular, massive stars, with their
strong winds and short lifetime have a strong impact on the
circumstellar environment. The high density of absorbing material
makes optical observations poorly suited for the observation of the
earliest stages of star formation, while the infrared (IR) bands $J,
H$ and $K$ have proven to be an excellent tool for unveiling embedded
star formation regions due to the reduced impact of dust extinction at
longer wavelengths. 

A role similar to IR observations is currently being played by X-ray
observations, to which dust is as transparent.  Given that X-ray
luminosity of stars evolves strongly with age, young stars are very
active X-ray sources, so that they can be easily selected from field
stars. X-ray surveys of young associations and clusters also provide
valuable information on the magnetic activity of young stars and on
their accretion rates, as well as effectively supplement other
approaches to membership determination in these regions (see e.g.
\citealp{fm99}, although the field is evolving rapidly). In comparison
with the low-mass domain X-ray observations of massive star-forming
regions and massive young stellar objects (YSOs) have been been rare,
with very few detections of X-ray emission from massive YSOs reported
to date (\citealp{kkh02}).

Recently \cite{lk02} (LDK02, hereafter) have published a study (based
on the 2MASS Point Source Catalogue -- PSC) of 22 young open clusters
in the Cygnus region, of which 12 objects are recently discovered
clusters and 3 are new clusters candidates. All of them appear to
contain a significant number of massive stars. We have searched the
\chandra\ and \xmm\ archives for observations of these regions, and
found a \chandra\ medium sensitivity observation (45 ks) in the
direction of one of the recently discovered clusters, which
corresponds to the well known region of star formation S106 (referred
to as ``Cl02'' in LDK02).  Note that there are no \chandra\ or \xmm\ 
observations to date in the direction of any of the other 21 young
clusters.

The H\,{\sc ii} region S106 is a prominent bipolar emission nebula
associated with an extended molecular cloud. The nebula appears to be
excited by a single massive star at its center, S106~IRS~4
(\citealp{ggc82}), also known as S106~IR. The star is surrounded by a
a compact region of radio emission indicating that this is a newly
formed star still embedded within its natal dust cocoon
(\citealp{sfm+83}; \citealp{cc03}). \cite{sld+82} have used
$UV\!B\!RI$ photometry and broadband polarimetry of field stars to
determine the foreground extinction and polarization, resulting in a
distance estimate for the molecular complex of $600 \pm 100$~pc
($DM=9$).

The S106 cluster was first studied by \cite{hr91} in the IR, who
counted $\sim 160$ cluster members of which half are within a radius
of 1.0 arcmin from the cluster centre -- which they position at about
30 arcsec west and 15 arcsec north of S106~IRS~4.  They derive a
cluster mass of $\sim 140\, M_{\sun}$ and estimate that star formation
in this cluster has been ongoing for the past 1--2 million years.

In their study, LDK02 position the cluster center at $\alpha,
\delta$=20:27:25, 37:22:48 (all coordinates are in the J2000 system),
which is about 21 arcsec west of the position of S106~IRS~4 in the
2MASS PSC (the coordinate of S106~IRS~4 being $\alpha,
\delta$=20:27:26.76, 37:22:47.9). From cumulative star counts they
derive a 90\% population radius $R_{90} = 1.1$ arcmin, even though
they caution that this value may only correspond to the core radius of
the structure.  They estimate the cluster to contain $20\pm 5$ OB star
and have a total mass in the range of 400--600$M_{\sun}$.
\cite{ssk+02} identify the cluster as the primary star formation site
in S106, coincident with the prominent peak in CO emission. They also
detected a second weaker CO peak located 5 arcmin south of S106~IRS~4,
which harbors a small IR cluster and nebulosity.  They interpret the
nebula as a signpost for a secondary site of star formation in S106
and refer to it as ``S106~south''.

In this article we present \chandra\ observations of the S106 and
S106~south clusters and report detection of S106~IRS~4 in the X-ray. The
paper is organized as follow: the observations and data analysis are
presented in Sect.~\ref{sec:obs}. The results are summarised in
Sect.~\ref{sec:analysis} and discussed in Sec.~\ref{sec:disc}.

\section{Observations}
\label{sec:obs}

The X-ray observations discussed in this paper were obtained with the
\chandra\ observatory. The 45 ks ACIS-I observation of S106 was taken
starting on November 3 2001 at 00:08:14 UT. The Principal Investigator
for these observations is Y. Maeda and the observation target is the
protostellar object S106~IRS~4. The data were retrieved from the public
data archive, with no re-processing done on the archival data.

We performed the source detection on the event list, using the Wavelet
Transform detection algorithm developed at Palermo Astronomical
Observatory (\textsc{Pwdetect}, available at
http://oapa.astropa.unipa.it/progetti\_ricerca/PWDetect).  The
threshold for source detection was taken as to ensure a maximum of one
spurious source per field.

For the brightest sources for which a spectral and timing analysis was
carried out source and background regions were defined in
\textsc{ds9}, and light curves and spectra were extracted from the
photon list using CIAO V.~2.2.1 threads, which were also used for the
generation of the relative response matrices. Spectral analysis was
performed in \textsc{xspec}.

\begin{figure*}[!tbp]
  \begin{center} \leavevmode 
        \epsfig{file=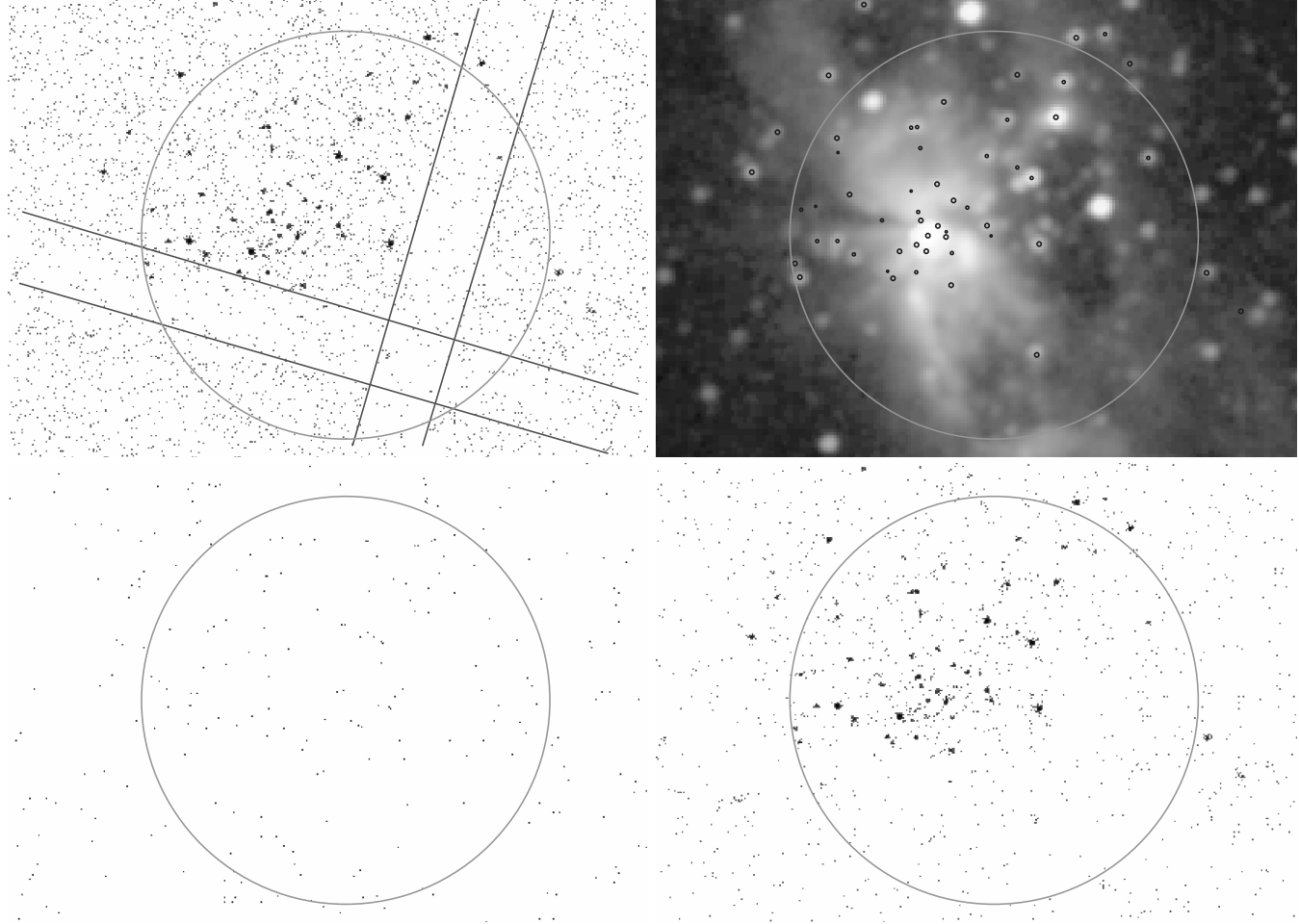, width=\hsize}

\caption{The field centered on S106 in the ACIS-I camera for the full
  energy range of the detectors (top left), in the 2MASS $K$-band (top
  right), in the ACIS-I camera for the energy range $\Delta E =
  0.2-1.0$~keV (bottom left) and in the ACIS-I camera for the energy
  range $\Delta E = 1.0-7.5$~keV (bottom right). The X-ray sources
  detected in the \chandra\ image have been overlaid on the 2MASS image
  and are represented by small circles. The core size of S106 is shown
  as a large circle (radius 1.1. arcmin). In the top left image the
  lines indicate the position of the CCD gaps in the ACIS camera.}

  \label{fig:image} \end{center}
\end{figure*}

Fig.~\ref{fig:image} shows the field centered on S106 as seen in the
ACIS-I camera (for various energy ranges) and in the $K$-band of the
2MASS survey\footnote{obtained from the 2MASS Quicklook Image Services
of the NASA/IPAC Infrared Science Archive.}. Only the area within 1.1
arcmin from the centre is displayed for clarity, although we applied
our analysis to an area within 4.4 arcmin from the cluster
centre. A radius of 4.4 arcmin was chosen with the aim of
studying the cluster with a minimal contamination from foreground and
background sources in the absence of a membership list. The chosen
radius corresponds to 4 times the S106 core radius as estimated by
LDK02 and therefore should contain the large majority of the cluster
members.  At the same time, it does not extends to regions too far
off-axis, where the sensitivity to X-ray sources would be lower
(because of vignetting and because the Chandra PSF widens).

A clustering of X-ray sources is visible in the ACIS-I image obtained
for the entire energy range ($\Delta E = 0.2-10.0$~keV), although the
cluster is 'cut' by the ACIS CCD gaps, which determine a reduced
sensitivity region. The 2MASS $K$-band image is dominated by the young
massive stellar object S106~IRS4 (a bright IR source) near the center
and by the double-lobed infrared nebula around it. S102~IRS4 is
thought to be the engine which powers the double-lobed nebula and the
associated H~{\sc ii} region (\citealp{ggc82}).  The X-ray sources
detected with \textsc{Pwdetect} are overlaid on the 2MASS image.

As discussed below, most X-ray sources (and in particular S106~IRS~4)
have an IR counterpart. A number of X-ray sources close to S106~IRS~4
have no counterpart unresolved from S106~IRS~4 in the 2MASS image,
while north of S106~IRS~4 an example of a double X-ray source with an
unresolved IR counterpart can be seen. The cluster center coordinates
as given by LDK02 (the center of the big circle in
Fig.~\ref{fig:image}) lies 0.3 arcmin off-axis in the \chandra\ 
observation.

In Fig.~\ref{fig:image} the ACIS-I images of the field centered on
S106 for a soft ($\Delta E = 0.2-1.0$~keV) and a hard energy band
($\Delta E = 1.0-7.5$~keV) are also shown. The great majority of the
X-ray sources in the direction of S106 are not visible for energies
lower the 1.0 keV, while are clearly visible in the hard band, as
expected on the basis of the high extinction toward the cluster (see
below).

In Table~\ref{tab:auto} we list the coordinates for all the X-ray
sources detected with \textsc{Pwdetect} within a radius of 4.4 arcmin
from the center of S106.  A total of 87 X-ray sources are detected, of
which 45 are within the S106 core radius. The X-ray source density
within the 4.4 arcmin radius is 1.4 arcmin$^{-2}$, 10 times higher
than in the rest of the ACIS-I field\footnote{excluding the sources
and the area of ``S106~south'' -- see next section.}. The X-ray source
density within the 1.1 arcmin core radius is 11.8~arcmin$^{-2}$. Most
of the detected sources are rather weak, with a count rate below 1 ct
ks$^{-1}$. The faintest source is source 9 with a count rate of
0.16~ks$^{-1}$, which corresponds\footnote{Assuming a plasma
temperature of $kT = 2.16$~keV, a metallicity of $Z=0.2~Z_{\sun}$ and
\nh\ as derived from IR data for this source. See Sect.~3.2.} to a
flux of $1.6 \times 10^{-15}$~\ecms and an intrinsic (unabsorbed) flux
of $4.2 \times 10^{-15}$~\ecms. Only 7 sources (sources 22, 30, 32
39, 60, 68 and 72) have a count rate just above 2~ct ks$^{-1}$.

For the sources with an IR counterpart the values of their magnitude
in the $J$, $H$ and $K$ bands are also given in Table~\ref{tab:auto}
together with the radial distance from the possible IR counterpart. An
X-ray source was considered to have a possible IR counterpart if an IR
source is present in the 2MASS PSC catalogue within a radius of 3
arcsec from the X-ray position.  Of the 87 sources within the 4.4
arcmin radius 64 have a possible IR counterpart. Source 82 has
two possible IR counterparts, while sources 55 and 58 have the same
IR counterpart.  All the 7 brightest X-ray sources have an IR
counterpart.

The X-ray counterpart of S106~IRS~4 is source 50 in our list. It is a
weak source with a count rate of $0.3\pm0.11$ ct~ks$^{-1}$; its X-ray
position is less than 0.5 arcsec away from the coordinates of
S106~IRS~4 in the 2MASS PSC, which agrees well with the coordinate for
S106~IRS~4 given in \cite{ssk+02} ($\alpha, \delta$= 20:27:26.74,
37:22:48), but are more than 12 arcsec away from the position given
for this object in the \textsc{Simbad} database, which corresponds to
the coordinates given in \cite{ggc82}.  

To quantify the visual impression that the great majority of the X-ray
sources detected in the direction of S106 have hard spectra, (and thus
are not visible below 1.0 keV) we have run \textsc{Pwdetect} also on
the energy-filtered ACIS-I event lists. Only 8 sources (flagged with a
's' in Table~\ref{tab:auto}) are detected by the algorithm in the soft
energy band ($\Delta E = 0.2-1.0$~keV) within a radius of 4.4 arcmin
from the center of S106. All of these 8 sources have an IR
counterpart, and only one is within the S106 core radius.  In the
total ACIS-I field 27 out 125 sources are detected in the soft band,
implying that, within 4.4 arcmin from the center of S106, the fraction
of soft sources (9\%) is more than 6 times lower then in the rest of
the ACIS-I field (60\%)\footnote{excluding the sources in
  ``S106~south'' -- see next section.}.

LDK02 derive, for the stars belonging to S106, extinction coefficients
in the $K$ band ranging from $A_K= 0.5$ to $A_K = 4.0$. Given the
conversion factor (\citealp{rl85}):
\begin{equation}
A_K = 0.11 A_V
\label{eq:akav}
\end{equation}
and the relation (see e.g.~\citealp{cox00}):
\begin{equation}
\nh/A_{V}=1.9\times 10^{21}$ ~cm$^{-2}$~mag$^{-1}
\label{eq:avnh}
\end{equation}
the absorbing column density for members of S106 must be $\nh \ga 8.5
\times 10^{21}$ ~cm$^{-2}$.  The fact that the majority of the sources
are not visible in the ACIS camera for energies below 1.0~keV is
consistent with such high absorbing column densities. Indeed, if
we assume as a typical spectrum for the sources listed in
Table~\ref{tab:auto} an absorbed Raymond-Smith plasma model with $\nh
= 2.0 \times 10^{21}$ ~cm$^{-2}$, plasma temperature $kT=1.5$~keV and
metal abundance $Z=0.2\, Z_{\sun}$ then, using the \textsc{pimms}
simulator at {\sc Heasarc}, the expected ratio of the count rate
between the soft and the total band is 0.15. This estimate is not
very sensitive to the assumed X-ray temperature in the range
1$-$~3~keV.

\begin{table*}[thbp]

  \caption{X-ray sources detected in the \chandra\ data within 4.4
    arcmin from the center of S106. For those sources for which an IR
    counterpart in the 2MASS PSC exists the $J, H, K$ magnitude are
    given together with the separation ($r$) from the possible
    counterparts. The 's' flag indicates sources which are also
    detected in the soft band ($\Delta E = 0.2-1.0$~keV). The
    value of $\log L_{X}$ derived from Eq.~3 for the sources with IR
    counterpart is also given. Seven sources are inconsistent with the
    estimated age and distance of S106, therefore, their luminosity
    cannot be computed using Eq. 3. We have marked these sources with
    'f'.}

  \begin{center}
    \leavevmode
    \begin{tabular}{r|cccrrrccc}
        \hline
        \hline
Source & RA(J2000) & Dec(J2000) & Count rate & $J$ & $H$ & $K$ & $r$ & $\log L_{X}$ & Soft \\
~      &          &             & (ks)$^{-1}$ & ~ & ~ & ~ & (arcsec) & ($\log~$\es) &~\\
\hline
  1 & 20:27:06.1 & 37:23:29 &   0.59 $\pm$ 0.18 &   14.5 &   13.3 &   13.0 &   0.68 & 29.6 & ~ \\ 
  2 & 20:27:07.4 & 37:21:11 &   0.40 $\pm$ 0.14 &   14.4 &   13.2 &   12.7 &   0.63 & 29.7 & ~ \\ 
  3 & 20:27:08.5 & 37:25:06 &   0.35 $\pm$ 0.12 &   14.9 &   13.3 &   12.5 &   1.05 & 29.9 & ~ \\ 
  4 & 20:27:09.0 & 37:22:18 &   0.37 $\pm$ 0.13 &   12.9 &   12.0 &   11.6 &   2.29 & 29.4 & ~ \\ 
  5 & 20:27:09.2 & 37:23:31 &   0.17 $\pm$ 0.09 &   13.3 &   12.3 &   11.9 &   0.40 & 29.2 & ~ \\ 
  6 & 20:27:09.2 & 37:24:32 &   0.55 $\pm$ 0.17 &   16.3 &   15.1 &   14.5 &   0.76 & 29.7 & ~ \\ 
  7 & 20:27:09.6 & 37:21:38 &   0.25 $\pm$ 0.18 &   13.6 &   12.6 &   12.2 &   0.97 & 29.3 & ~ \\ 
  8 & 20:27:09.9 & 37:22:28 &   0.57 $\pm$ 0.18 &   16.1 &   14.4 &   13.4 &   0.44 & 30.1 & ~ \\ 
  9 & 20:27:10.7 & 37:24:42 &   0.16 $\pm$ 0.07 &   15.8 &   14.5 &   13.9 &   0.28 & 29.2 & ~ \\ 
 10 & 20:27:12.1 & 37:23:60 &   0.44 $\pm$ 0.15 &   15.0 &   13.1 &   12.0 &   0.28 & 30.2 & ~ \\ 
 11 & 20:27:12.2 & 37:22:53 &   0.18 $\pm$ 0.08 &   15.5 &   14.1 &   13.6 &   0.54 & 29.3 & ~ \\ 
 12 & 20:27:12.5 & 37:25:02 &   0.48 $\pm$ 0.16 &   16.1 &   14.5 &   13.9 &   0.63 & 29.9 & ~ \\ 
 13 & 20:27:13.1 & 37:22:07 &   0.16 $\pm$ 0.08 &   17.2 &   15.2 &   14.4 &   0.45 & 29.6 & ~ \\ 
 14 & 20:27:13.3 & 37:22:13 &   0.49 $\pm$ 0.15 &    - &    - &    - &   - &  - & ~ \\ 
 15 & 20:27:14.9 & 37:22:43 &   0.21 $\pm$ 0.08 &   16.0 &   14.2 &   13.6 &   0.35 & 29.6  & ~ \\ 
 16 & 20:27:16.4 & 37:22:02 &   0.28 $\pm$ 0.12 &   15.5 &   14.8 &   14.6 &   0.47 & f & ~ \\ 
 17 & 20:27:18.0 & 37:24:39 &   1.29 $\pm$ 0.19 &   16.0 &   14.0 &   13.2 &   0.42 & 30.6 & ~ \\ 
 18 & 20:27:18.3 & 37:22:23 &   0.34 $\pm$ 0.14 &    - &    - &    - &   - &  - & ~ \\ 
 19 & 20:27:18.4 & 37:21:04 &   0.20 $\pm$ 0.09 &    - &    - &    - &   - &  - & ~ \\ 
 20 & 20:27:19.2 & 37:22:36 &   0.72 $\pm$ 0.21 &   15.0 &   13.5 &   12.9 &   0.65 & 30.1  & ~ \\ 
 21 & 20:27:20.8 & 37:23:13 &   0.23 $\pm$ 0.11 &   15.1 &   13.3 &   12.4 &   0.37 & 29.8 & ~ \\ 
 22 & 20:27:21.3 & 37:23:43 &   3.53 $\pm$ 0.46 &   17.4 &   15.1 &   13.8 &   0.18 & 31.1 & ~ \\ 
 23 & 20:27:21.4 & 37:24:28 &   0.23 $\pm$ 0.09 &   16.8 &   14.9 &   13.7 &   0.64 & 29.8 & ~ \\ 
 24 & 20:27:22.0 & 37:23:53 &   0.18 $\pm$ 0.08 &   16.0 &   13.8 &   12.5 &   0.25 & 29.8 & ~ \\ 
 25 & 20:27:22.5 & 37:19:50 &   0.78 $\pm$ 0.15 &    - &    - &    - &   - &  - & ~ \\ 
 26 & 20:27:22.8 & 37:23:52 &   1.48 $\pm$ 0.20 &   13.4 &   12.1 &   11.6 &   0.22 & 30.3 & ~ \\ 
 27 & 20:27:22.8 & 37:24:14 &   1.05 $\pm$ 0.17 &   14.2 &   13.6 &   13.3 &   0.09 & f & s \\ 
 28 & 20:27:23.1 & 37:23:38 &   0.29 $\pm$ 0.11 &   13.8 &   12.2 &   11.1 &   0.15 & 29.9 & ~ \\ 
 29 & 20:27:23.3 & 37:23:26 &   0.83 $\pm$ 0.16 &   13.8 &   11.1 &    9.0 &   0.45 & 30.8 & ~ \\ 
 30 & 20:27:23.8 & 37:22:45 &   2.12 $\pm$ 0.23 &   13.3 &   11.9 &   11.6 &   0.28 & 30.5 & s \\ 
 31 & 20:27:23.8 & 37:22:09 &   0.70 $\pm$ 0.31 &   14.1 &   12.8 &   11.9 &   1.22 & 30.2 & ~ \\ 
 32 & 20:27:24.0 & 37:23:07 &   2.31 $\pm$ 0.24 &   11.3 &   10.6 &   10.2 &   0.23 & 30.2 & ~ \\ 
 33 & 20:27:24.4 & 37:23:10 &   0.49 $\pm$ 0.16 &   15.2 &   13.5 &   11.5 &   0.38 & 30.3 & ~ \\ 
 34 & 20:27:24.4 & 37:23:40 &   0.32 $\pm$ 0.12 &   15.4 &   14.0 &   13.6 &   0.22 & 29.5 & ~ \\ 
 35 & 20:27:24.6 & 37:18:29 &   0.67 $\pm$ 0.19 &   13.4 &   13.0 &   12.9 &   0.37 & f & s \\ 
 36 & 20:27:24.6 & 37:23:25 &   0.32 $\pm$ 0.12 &   14.0 &   12.5 &   12.0 &   0.08 & 29.7 & ~ \\ 
 37 & 20:27:25.1 & 37:22:48 &   0.32 $\pm$ 0.13 &    - &    - &    - &   - &  - & ~ \\ 
 38 & 20:27:25.2 & 37:22:51 &   0.65 $\pm$ 0.19 &   14.0 &   13.0 &   12.9 &   0.28 & 28.8 & ~ \\ 
 39 & 20:27:25.2 & 37:23:14 &   2.18 $\pm$ 0.24 &   14.3 &   12.3 &   11.2 &   0.10 & 30.9 & ~ \\ 
 40 & 20:27:25.7 & 37:22:57 &   0.40 $\pm$ 0.13 &    - &    - &    - &   - &  - & ~ \\ 
 41 & 20:27:26.1 & 37:22:59 &   0.41 $\pm$ 0.13 &   13.0 &   12.0 &   11.0 &   2.06 & 29.9 & ~ \\ 
 42 & 20:27:26.1 & 37:22:42 &   0.17 $\pm$ 0.08 &    - &    - &    - &   - &  - & ~ \\ 
 43 & 20:27:26.2 & 37:22:32 &   0.56 $\pm$ 0.17 &   13.5 &   11.1 &   10.3 &   0.74 &  30.4& ~ \\ 
 44 & 20:27:26.3 & 37:22:49 &   0.19 $\pm$ 0.11 &    - &    - &    - &   - &  - & ~ \\ 
 45 & 20:27:26.3 & 37:22:47 &   0.76 $\pm$ 0.21 &    - &    - &    - &   - &  - & ~ \\ 
 46 & 20:27:26.3 & 37:23:31 &   0.26 $\pm$ 0.10 &   15.1 &   13.2 &   12.2 &   0.61 & 29.9 & ~ \\ 
 47 & 20:27:26.5 & 37:24:30 &   0.41 $\pm$ 0.13 &   15.7 &   14.1 &   13.6 &   0.57 & 29.8 & ~ \\ 
 48 & 20:27:26.5 & 37:22:51 &   0.34 $\pm$ 0.12 &    - &    - &    - &   - &  - & ~ \\ 
 49 & 20:27:26.5 & 37:23:05 &   0.36 $\pm$ 0.14 &   12.3 &   10.8 &   11.0 &   1.19 & 29.5 & ~ \\ 
 50 & 20:27:26.8 & 37:22:48 &   0.30 $\pm$ 0.11 &   10.4 &    7.7 &    5.9 &   0.44 & 30.3 & ~ \\ 

\end{tabular}
    \label{tab:auto}
  \end{center}
\end{table*}

\addtocounter{table}{-1}
\begin{table*}[!thbp]
  \begin{center}
    \caption{{\em (continued)} X-ray sources within 4.4 arcmin of S106 center.}
    \leavevmode
        \begin{tabular}{r|cccrrrccc}
Source & RA(J2000) & Dec(J2000) & Count rate & $J$ & $H$ & $K$ & $r$ & $\log L_{X}$ & Soft\\
~      &           &             & (ks)$^{-1}$ & ~ & ~ & ~ & (arcsec)& ($\log~$\es) &~\\
\hline

 51 & 20:27:26.8 & 37:22:43 &   0.30 $\pm$ 0.11 &    - &    - &    - &   - &  - & ~ \\ 
 52 & 20:27:27.0 & 37:22:53 &   0.23 $\pm$ 0.09 &    - &    - &    - &   - &  - & ~ \\ 
 53 & 20:27:27.0 & 37:23:16 &   0.27 $\pm$ 0.11 &    - &    - &    - &   - &  - & ~ \\ 
 54 & 20:27:27.1 & 37:22:55 &   0.98 $\pm$ 0.16 &   12.2 &   11.9 &   10.5 &   2.08 & 30.2 & ~ \\ 
 55 & 20:27:27.1 & 37:23:23 &   0.43 $\pm$ 0.14 &   13.9 &   12.3 &   11.6 &   0.49 &  30.0 & ~ \\ 
 56 & 20:27:27.1 & 37:22:45 &   0.19 $\pm$ 0.08 &    - &    - &    - &   - &  - & ~ \\ 
 57 & 20:27:27.1 & 37:22:36 &   0.53 $\pm$ 0.16 &    - &    - &    - &   - &  - & ~ \\ 
 58 & 20:27:27.2 & 37:23:23 &   0.39 $\pm$ 0.13 &   13.9 &   12.3 &   11.6 &   1.54 & 29.9 & ~ \\ 
 59 & 20:27:27.2 & 37:23:02 &   0.26 $\pm$ 0.11 &   13.5 &   11.0 &    9.8 &   2.18 & 30.1 & ~ \\ 
 60 & 20:27:27.6 & 37:22:43 &   2.45 $\pm$ 0.25 &   14.0 &   12.6 &    10.8 &  3.02 &  31.0 & ~ \\ 
 61 & 20:27:27.7 & 37:22:34 &   0.35 $\pm$ 0.13 &    - &    - &    - &   - &  - & ~ \\ 
 62 & 20:27:27.9 & 37:22:36 &   0.42 $\pm$ 0.14 &   14.6 &   13.1 &    9.5 &   2.12 & 30.5 & ~ \\ 
 63 & 20:27:28.0 & 37:21:29 &   0.21 $\pm$ 0.08 &    - &    - &    - &   - &  - & ~ \\ 
 64 & 20:27:28.0 & 37:22:53 &   0.43 $\pm$ 0.15 &    - &    - &    - &   - &  - & ~ \\ 
 65 & 20:27:28.2 & 37:24:47 &   0.96 $\pm$ 0.15 &   16.4 &   14.6 &   13.7 &   0.16 &  30.3 & ~ \\ 
 66 & 20:27:28.5 & 37:24:03 &   0.28 $\pm$ 0.10 &   14.8 &   13.2 &   12.6 &   0.35 &  29.7 & ~ \\ 
 67 & 20:27:28.8 & 37:19:50 &   0.68 $\pm$ 0.20 &   13.4 &   12.8 &   12.6 &   0.21 & f & s \\ 
 68 & 20:27:28.8 & 37:19:45 &   3.71 $\pm$ 0.30 &   13.7 &   13.2 &   12.8 &   0.39 & f & s \\ 
 69 & 20:27:28.8 & 37:22:42 &   0.44 $\pm$ 0.15 &   15.0 &   13.4 &   12.8 &   0.71 & 29.9 & ~ \\ 
 70 & 20:27:28.9 & 37:23:01 &   0.62 $\pm$ 0.11 &    - &    - &    - &   - &  - & ~ \\ 
 71 & 20:27:29.2 & 37:23:15 &   0.27 $\pm$ 0.10 &    - &    - &    - &   - &  - & ~ \\ 
 72 & 20:27:29.2 & 37:22:46 &   2.48 $\pm$ 0.25 &   14.7 &   12.9 &   12.1 &   0.10 &  30.8 & ~ \\ 
 73 & 20:27:29.3 & 37:23:19 &   0.20 $\pm$ 0.08 &   15.4 &   13.9 &   12.6 &   0.50 &  29.8 & ~ \\ 
 74 & 20:27:29.5 & 37:23:40 &   0.98 $\pm$ 0.16 &   14.9 &   13.1 &   12.2 &   0.25 &  30.4 & ~ \\ 
 75 & 20:27:29.8 & 37:22:46 &   0.28 $\pm$ 0.10 &   15.6 &   13.9 &   13.0 &   0.29 &  29.8 & ~ \\ 
 76 & 20:27:29.8 & 37:22:57 &   0.17 $\pm$ 0.09 &    - &    - &    - &   - &  - & ~ \\ 
 77 & 20:27:30.2 & 37:22:56 &   0.30 $\pm$ 0.11 &    - &    - &    - &   - &  - & ~ \\ 
 78 & 20:27:30.3 & 37:22:34 &   0.71 $\pm$ 0.25 &   14.8 &   13.1 &   12.2 &   0.31 &  30.3 & ~ \\ 
 79 & 20:27:30.4 & 37:22:39 &   0.34 $\pm$ 0.14 &   15.5 &   13.9 &   13.5 &   0.21 &  29.6 & ~ \\ 
 80 & 20:27:30.7 & 37:24:29 &   0.31 $\pm$ 0.11 &   12.9 &   12.4 &   12.2 &   0.22 & f & s \\ 
 81 & 20:27:30.8 & 37:24:11 &   0.18 $\pm$ 0.08 &   16.3 &   14.3 &   13.2 &   0.57 &  29.8 & ~ \\ 
82{\tiny-a} &  20:27:30.9   &   37:23:21  &   0.34 $\pm$ 0.12 &   15.5 &   14.0 &   14.3 &   0.21 &  29.6 & ~ \\ 
82{\tiny-b} & \tiny{idem}  &  \tiny{idem} &   \tiny{idem} &   15.6 &   14.0 &   13.7 &   2.63 &  28.5 & ~ \\
 83 & 20:27:31.6 & 37:23:08 &  0.67 $\pm$ 0.13 &   14.5 &   12.9 &   12.2 &   0.17 &  30.2 & ~ \\ 
 84 & 20:27:35.5 & 37:20:24 &  0.36 $\pm$ 0.13 &   11.7 &   11.2 &   11.1 &   2.58 & f & s \\ 
 85 & 20:27:40.5 & 37:21:47 &  0.27 $\pm$ 0.11 &   16.2 &   15.1 &   14.6 &   0.41 & 29.2 & ~ \\ 
 86 & 20:27:41.6 & 37:25:35 &  0.29 $\pm$ 0.12 &    - &    - &    - &   - &  - & ~ \\ 
 87 & 20:27:42.7 & 37:23:40 &  1.06 $\pm$ 0.17 &   14.0 &   13.2 &   12.8 &   0.46 & 29.2 & s \\

\hline
   \end{tabular}
  \end{center}
\end{table*}

\begin{figure*}[!htbp]
  \begin{center} \leavevmode 
        \epsfig{file=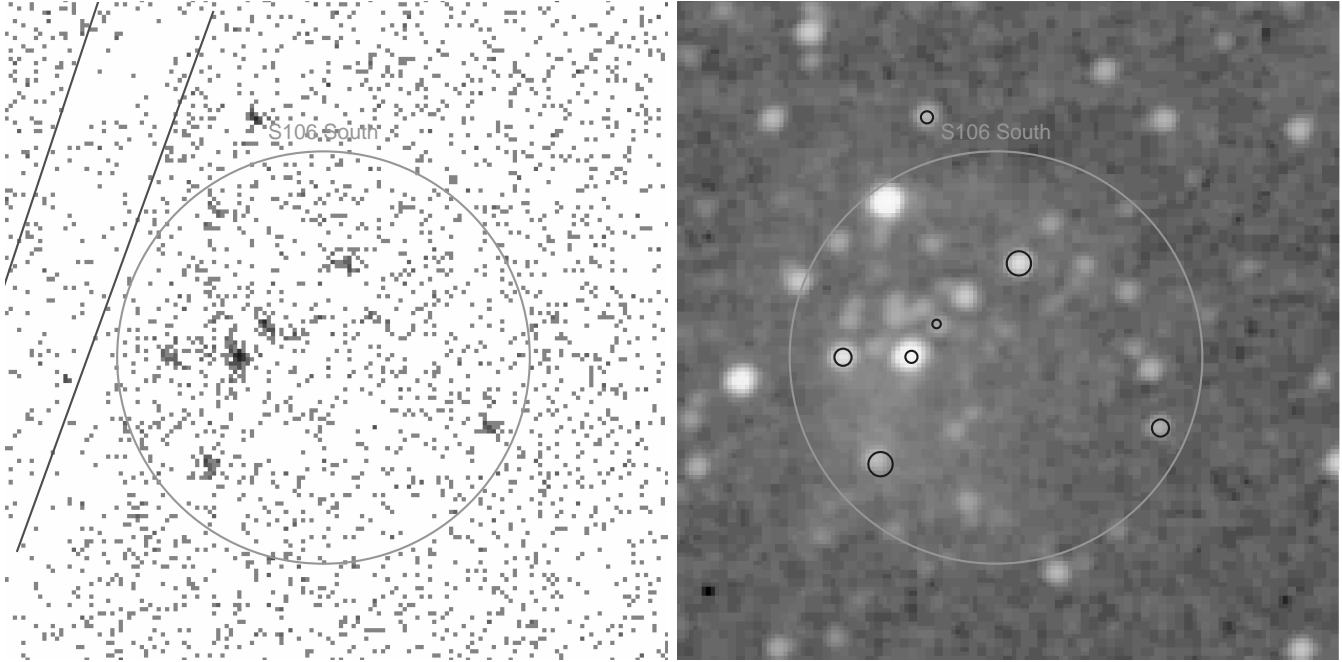, width=\hsize}

\caption{The field centered on S106~south in the ACIS-I camera (left)
  and in the 2MASS $K$-band image (right). The X-ray sources detected
  in the \chandra\ image have been overlaid on the 2MASS image and are
  represented by small circles with size proportional to the X-ray
  count rate. The selection area that we have used for S106 south is
  shown as a large circle (radius 0.8 arcmin). In the ACIS image the
  lines indicate the position of the CCD gaps.}

  \label{fig:S106south} \end{center}
\end{figure*}

\subsection{S106~south}

As mentioned in the Introduction, \cite{ssk+02} identify the cluster
S106 as a primary star formation site coincident with the prominent
peak in CO emission. They also detect a second weaker CO peak located
5 arcmin south of S106~IRS~4 which harbors a small IR cluster and
nebulosity and refer to it as ``S106~south''.

Fig.~\ref{fig:S106south} shows the field centered around S106~south in
the ACIS-camera and in the 2MASS $K$ band. An increase in source
density in the direction of S106~south is visible in the \chandra\
image. The approximate center of this group of X-ray sources is at
$\alpha, \delta =$~20:27:23.2, 37:17:33 (5 arcmin off-axis in the
\chandra\ data). In Table~\ref{tab:autosouth} the coordinates and
2MASS IR photometry for all X-ray sources within a radius of $0.8$
arcmin from this position are listed.  A total of 6 sources have been
detected, making the source density within this area 20 times higher
than in the rest of the field (excluding the area of $4.4$ arcmin
radius from the center of S106).

All of the 6 sources have an IR counterpart within a radial distance
of $3$ arcsec, and only one of them is visible in the soft band.

\begin{table*}[!thbp]
  \begin{center}

  \caption{X-ray sources detected in the \chandra\ data using {\sc
      Pwdetect} within 0.8 arcmin from the center of S106 south. For
      those sources for which an IR counterpart in the 2MASS PSC
      exists the $J, H, K$ magnitude are given. The 's' flag indicates
      sources which are also detected in the soft band ($\Delta E =
      0.2-1.0$~keV). The value of $\log L_{X}$ derived from Eq.~3
      is also given.}


    \leavevmode
    \begin{tabular}{r|cccrrrccc}
        \hline
        \hline
Source & RA(J2000) & Dec(J2000) & Count rate & $J$ & $H$ & $K$ & $r$ & $\log L_{X}$ & Soft \\
~      &          &             & (ks)$^{-1}$ & ~ & ~ & ~ & (arcsec) & ($\log~$\es) & ~\\
\hline
 1 & 20:27:20.0 & 37:17:16 &  0.51 $\pm$ 0.17 &   14.6 &   13.5 &   13.1 &   0.57 & 29.5 & ~ \\ 
 2 & 20:27:22.8 & 37:17:55 &  0.64 $\pm$ 0.19 &   15.2 &   13.2 &   11.6 &   0.27 & 30.4 & ~ \\ 
 3 & 20:27:24.4 & 37:17:41 &  0.38 $\pm$ 0.15 &   15.8 &   14.1 &   13.6 &   0.79 & 29.8 & ~ \\ 
 4 & 20:27:24.8 & 37:17:33 &  1.28 $\pm$ 0.20 &   11.0 &   10.2 &    9.8 &   0.58 & 30.1 & ~ \\ 
 5 & 20:27:25.4 & 37:17:08 &  0.82 $\pm$ 0.23 &   16.5 &   14.3 &   12.6 &   1.71 & 30.6 & s \\ 
 6 & 20:27:26.2 & 37:17:33 &  0.50 $\pm$ 0.17 &   14.8 &   12.7 &   11.4 &   0.44 & 30.3 & ~ \\ 
\hline  
   \end{tabular}
  \label{tab:autosouth}
  \end{center}
\end{table*}

\section{Results}
\label{sec:analysis}

\subsection{Color-magnitude and color-color diagrams for the IR
  counterparts to X-ray sources}

\begin{figure}[!tbp]
  \begin{center} \
        \leavevmode 
        \epsfig{file=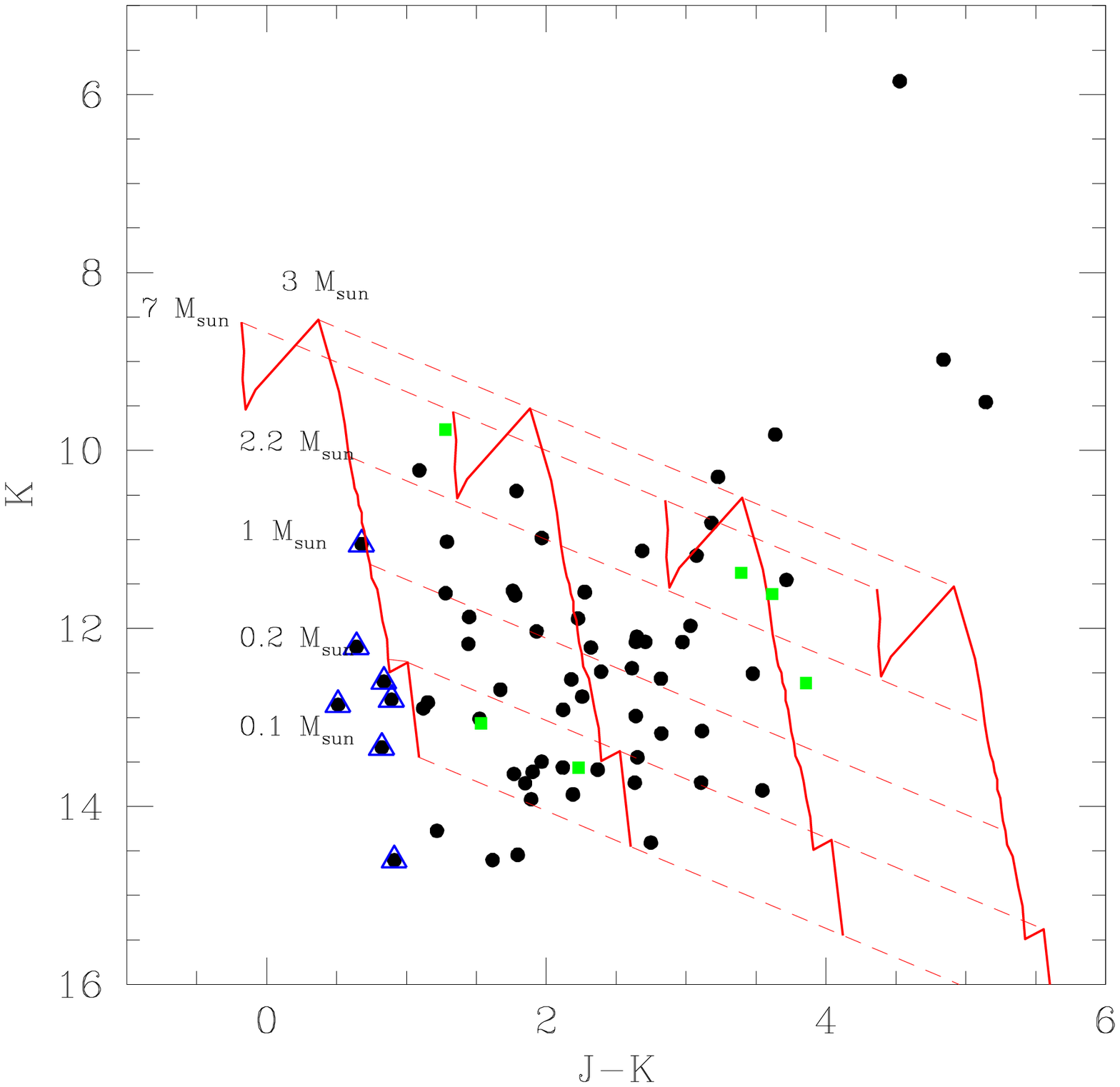, width=\hsize}

        \caption{Color magnitude diagram for the IR counterparts to
          the X-ray sources detected within 4.4 arcmin from the center
          of S106 (black circles) and within S106~south (gray
          squares).  Theoretical 2 Myr isochrones are also shown for
          different extinction coefficients (from left to right
          $A_K={\rm 0, 1, 2, 3}$).  The dashed lines indicate the
          reddening vector for stars of constant mass.  Points with a
          triangle indicate sources inconsistent with the theoretical
          2 Myr isochrone (see text for details).}
  \label{fig:cmd_2myr}
  \end{center}
\end{figure}

To compare the population of the X-ray sources detected in the
\chandra\ exposure with the population of S106 as defined by LDK02 we
have constructed a $K$ vs. $J-K$ color magnitude diagram
(Fig.~\ref{fig:cmd_2myr}) for the X-ray sources listed in
Tables~\ref{tab:auto} and~\ref{tab:autosouth} which have an IR
counterpart in the 2MASS PSC.

In the diagram 2 Myr isochrones are shown for 4 values of extinction
($A_K={\rm 0, 1, 2, 3}$). Reddening vectors have been computed
following the same relation adopted by LDK02, $A_K = R_K \times
E(J-K)$ where $E(J-K)$ is the colour excess ($E(J-K) = (J-K) -
(J-K)_0$) and $R_K = 0.66$ (\citealp{rl85}). The isochrones are from
\cite{sdf00}, for a metal abundance $Z=0.02$ plus overshooting,
shifted to the estimated distance of S106 (600 pc). Various
simplifying assumptions apply to these models, e.g. they include
neither rotation nor accretion, and we caution that evolutionary
models for pre-main sequence stars are not yet well
established. \cite{smv+04} recently reported the discovery of a
double-lined, spectroscopic, eclipsing binary in the Orion
star-forming region, with measured masses of 1.01~M$_{\sun}$ for the
primary and 0.73~M$_{\sun}$ for the secondary. They used their
measurements of the fundamental stellar properties for both components
to test the predictions of pre-main sequence stellar evolutionary
tracks. None of the models they examined (including the one by
\citealp{sdf00}) correctly predicts the masses of the two components
simultaneously.

In Fig.~\ref{fig:cmd_2myr} seven sources have been highlighted
using a triangle (sources 16, 27, 35, 67, 68, 80 and 84). These
sources fall on the blue side of the 2 Myr isochrone for zero
extinction and therefore are inconsistent with the estimated age and
distance of S106. There is evidence that these maybe foreground
sources. They do not meet the membership criteria used by LDK02 (their
extinction coefficient being too low) and
all of them but source 16 are detected in the \chandra\ soft band,
consistent with their being less absorbed than the others.  Note,
however, that sources  30 and 87, that have been flagged as
``soft'', are not inconsistent with a 2 Myr isochrone and the value of
their extinction coefficient as computed by LDK02 is consistent with
their membership criteria.

The sources in our sample appear very scattered in the color-magnitude
diagram, implying highly variable absorption. LDK02 find the same
large scatter for their IR sample of sources belonging to S106
(according to their selection criteria), and suggest that, while this
scatter is partly due to large and variable absorption, it is probably
also an effect of the unreliability of the 2MASS photometric data in
this very crowded area\footnote{The 2MASS images resolution is 1
arc-sec/pixel.}. From the color-magnitude diagram it appears that the
X-ray selected sample contains a number of heavily absorbed massive
stars. The brightest of all, with $K=5.9$, is the massive young
stellar object S106~IRS~4 at the center of the S106 complex.

The color-magnitude diagram also provides an indication of the
completeness limit of our X-ray survey, which is at around $K \simeq
12$, as we have verified by plotting differential star counts versus
apparent magnitude. For a 2 Myr isochrone $K = 12$ corresponds to
stars of mass $M \simeq 0.5\,M_{\sun}$ for low extinction ($A_K<0.1$)
and $M \ga 1.2\,M_{\sun}$ for extinction coefficients $A_K\ge 1.0$

\begin{figure}[!tbp]
  \begin{center} \
        \leavevmode 
        \epsfig{file=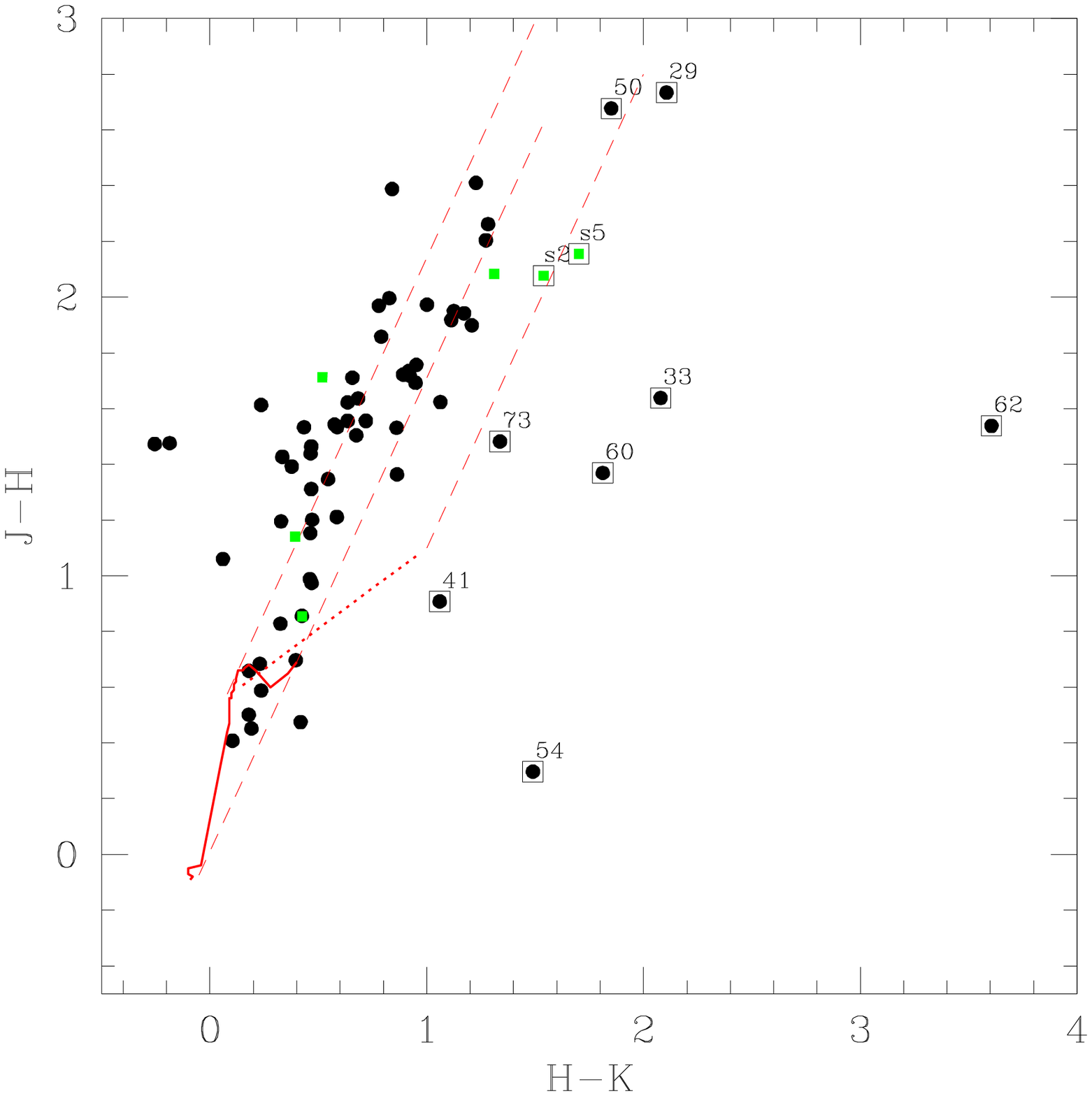, width=\hsize}

        \caption{Color-color diagram for the IR counterparts to the
          X-ray sources detected within 4.4 arcmin from the center of
          S106 (circular black points) and within S106~south (square
          gray points).  The solid line at lower-left is the
          theoretical 2 Myr isochrone.  The dotted line yields
          the locus of dereddened colours of classical T Tauri stars
          according to Meyer et al. (1997). The dashed lines mark the
          reddening bands. The most heavily reddened stars are
          identified.}

  \label{fig:ccd_2myr}
  \end{center}
\end{figure}

The $J-H$ versus $H-K$ color-color diagram in Fig.~\ref{fig:ccd_2myr}
allows normally reddened stars to be discriminated from star with IR
excess (indicative of warm circumstellar dust in addition to a
reddened photosphere). The intrinsic colours for stars on a 2 Myr
isochrone are indicated by the solid line, whereas the dotted line
yields the locus of dereddened colours of classical T Tauri stars
according to \cite{mch97}. The dashed lines define the region of
normal reddening using the reddening law from \cite{rl85} (the normal
reddening region of main sequence stars would be very similar to the
one of the stars on the 2 Myr isochrone).  The scatter of the X-ray
sources outside this region is significant and it is probably an
effect of the unreliability of the 2MASS photometric data in this
crowded area.  Nevertheless in the diagram we have identified 10
sources which appear to have significant IR excess (i.e. lie to the
right of the normal reddening region) and therefore could be embedded
young objects.  The presence of an IR excess for source 50
(S106~IRS~4) is already known from the literature (\citealp{fms+84};
\citealp{ggc82}; \citealp{hr91}).

In the color-magnitude diagram sources 29, 50 (S106~IRS~4) and 62
appear to be massive objects (with masses above $7\, M_{\sun}$).  This
estimate does not take into account the presence of an IR excess which
may lead to overestimate the star mass. This is not the case however
for S106~IRS~4, for which estimates in the literature independently
indicate a mass greater than $15\,M_{\sun}$ (\citealp{fms+84}).

\subsection{X-ray luminosity as a function of stellar mass} 
\label{sec:xrl}


To convert detector count rates to intrinsic X-ray luminosity we have
derived, using the {\sc pimms} software at {\sc Heasarc}, a conversion
factor between (unabsorbed) source flux and observed count rate,
taking into account the absorption. The absorbing column density for
each star was derived by first estimating $A_K$ and then deriving \nh\ 
using Eqs.~(\ref{eq:akav}) and~(\ref{eq:avnh}). The value of $A_K$ was
determined assuming that all sources in S106 and S106~south would
intrinsically lie on the theoretical 2 Myr isochrone and then
measuring the displacement along the reddening vector. No extinction
can obviously be derived for sources 16, 27, 35, 67, 68, 80 and 84,
which lie on the blue side of the isochrone. 

Because of the shape of the 2 Myr isochrone a degeneracy is present
for mass ranges 2.2--7.0\,$M_{\sun}$ and 0.2--0.3\,$M_{\sun}$; lacking
additional information (e.g. spectral types or optical magnitudes)
this degeneracy cannot be resolved. We thus, somewhat arbitrarily,
decided to associate the mass derived through the rightmost
intersection with the isochrone for the mass range
2.2--7.0\,$M_{\sun}$, and redefined the isochrone for the mass range
0.2--0.3\,$M_{\sun}$ using a tangent along the reddening vector at
$M=0.2\,M_{\sun}$ (shown in Fig.~\ref{fig:cmd_2myr} as a thin line).


 
As a typical X-ray spectrum for our sample we assumed a
one-temperature plasma model with $kT = 2.16$~keV\footnote{The value
is the same as the one assumed by \cite{fdm+03} for their analysis of
stars in the ONC, to which we will be comparing our results; this is
similar to the average plasma temperature of $kT=2.4$~keV derived
for the brightest sources in Table~3 (next section), excluding source
30 (because of the large uncertainty) and source 68 (because likely a
foreground source)} and a coronal metal abundance $Z=0.2 Z_{\sun}$,
deriving, for $\nh=1.0 \times 10^{22}$~cm$^{-2}$ and a count rate of
1~ks$^{-1}$, a conversion factor of $F_{\rm X~ intr} = 2.7 \times
10^{-14}$\ecms. The conversion factor scales almost linearly with
$\nh$ (for $\nh=[0.5-6.0]\times 10^{22}$~cm$^{-2}$, the range in which
all our objects fall, see below), so that
\begin{equation}
F_{\rm X~ intr} = 2.7 \times 10^{-14} \times \nh\ \times {\rm
ct}~~~[{\rm erg~cm^{-2}~s^{-1}}]
\label{eq:cf}
\end{equation}
where ${\rm ct}$ are the source counts in units of ks$^{-1}$ and
\nh\ is the absorbing column density in units of
$10^{22}$~cm$^{-2}$. The error introduced in the flux estimate by the
linear approximation is below 30\%.

In all but three cases we derived a value of $A_K$ corresponding to an
absorbing column density within the range $[0.5-6.0] \times
10^{22}$~cm$^{-2}$. For sources 38, 82-a and 87 we derived values of
$A_K \le 0.07 $ (corresponding to $\nh <0.1~\times
10^{22}$~cm$^{-2}$), significantly below the absorbing column
densities derived for all the other sources in our sample (which has a
median of $A_K=1.1$); we have therefore excluded these sources from
the rest of the analysis. Having determined for a star its intrinsic
magnitude in $K$, as described above, an estimate of its mass is
readily available by using the 2 Myr isochrone.

\begin{figure}[!tbp]
  \begin{center} \
        \leavevmode 
        \epsfig{file=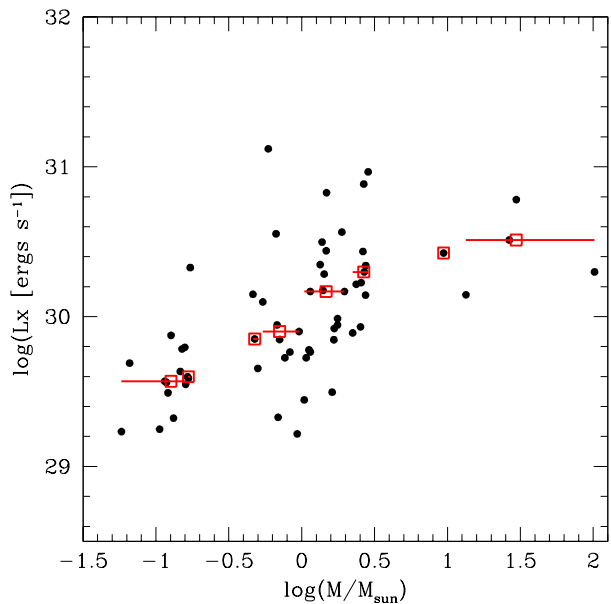, width=\hsize}

\caption{Scatter plot of intrinsic X-ray luminosities as a function of
  mass for the sample of 61 sources defined in the text. The gray
  horizontal lines indicate the median $L_{\rm X}$ for the
  corresponding mass interval. The abscissa of the empty-squares is
  the median mass for each subsample. The same mass intervals (in
  units of $M_{\sun}$) as in \cite{fdm+03} have been considered:
  $(0,0.16]$, $(0.16,0.25]$, $(0.25,0.5]$, $(0.5,1.0]$, $(1.0,2.0]$,
  $(2.0,3.0]$, $(3.0,10]$ and $(10, \infty)$}.

  \label{fig:LxM}
  \end{center}
\end{figure}

The derived X-ray luminosity as a function of the stellar mass is
shown in Fig.~\ref{fig:LxM}; 61 sources are included in the final
samples, i.e. all sources in Tables~\ref{tab:auto} and
\ref{tab:autosouth} with the exception of sources 16, 27, 35, 38, 67,
68, 80, 82-a, 84 and 87. The median X-ray luminosity for stars
in different mass intervals is also shown, using the same mass bins
adopted by \cite{fdm+03} in their analysis of the X-ray activity
indicators for members of the Orion Nebula Cluster (ONC). The median
X-ray luminosities are summarised in Table~\ref{tab:lx}. We note
that for masses lower than $0.5\,M_\odot$ the derived median X-ray
luminosities are systematically higher than the corresponding ONC
stars while the reverse is true for $M > 0.5\,M_\odot$. The values at
lower masses ($M \le 0.5\,M_\odot$), which corresponds to lower
fluxes, are likely biased to higher X-ray luminosities due to lack of
a parent sample of cluster members. In their analysis of the ONC
\cite{fdm+03} made use of an optical sample of cluster members in
order to place upper limits on X-ray luminosities. Lacking such a
sample for the S106 clusters we are not able to place upper
limits. The fact that for $M > 0.5\,M_\odot$ the values of median
$L_X$ for our sample are systematically lower than for the ONC could
be an indication of some physical difference between the S106 region
and the ONC, such as a different fraction of binaries. Nevertheless
the statistic is too small and the uncertainties too large to attach
any significance to this difference.

A trend of increasing X-ray luminosities with star mass is clearly
visible in Fig.~\ref{fig:LxM}. Two areas, one for
0.17--0.46~$M_{\sun}$ and one for 2.9--9.4~$M_{\sun}$, are devoid of
points. This is, partly, an effect of the way we have chosen to
circumvent the degeneracy in value of $A_K$ for some mass ranges (see
above), partly, it is a feature of the data themselves.  We note
also that lacking optical/IR spectral studies of the members of the
S106 clusters we do not know neither the spectral energy distribution
nor the spectral characteristics of the stars in our sample and are
therefore unable to classify them.

In Fig.~\ref{fig:LxM} the star with a mass estimate of
$102$~$M_{\sun}$ is source 50, corresponding to the central exciting
star S106~IRS~4. The value of $102$~$M_{\sun}$ is very probably
overestimated, due to the fact that the method we have used to derive
a mass value for each star does not take into account the possible
presence of an IR excess, which in this case is known to be present
(e.g. \citealp{fms+84} and Fig.~\ref{fig:ccd_2myr}). Indeed, the
masses of all stars in our sample for which we identified a
significant IR excess maybe affected by such an error.  For S106~IRS~4
the estimates in the literature indicate a mass greater than
$15\,M_{\sun}$ (\citealp{fms+84}) and therefore the presence of an IR
excess does not affect the placement of S106~IRS~4 in the highest mass
bin.

\begin{table}[thbp]
  
   \caption{Median value of the X-ray luminosity (second column) for
   our sample of 61 sources in S106 and S106~south for different mass
   ranges. These values can be compared with the values for members of
   the ONC (fourth column), as derived by \cite{fdm+03}. The number of
   sources of our sample within each mass range is also given.}

\begin{center} 
\leavevmode
    \begin{tabular}{ccrc}

\hline
\hline
Mass  & $\log(L_{\rm X})$ & N. src.  & $\log(L_{\rm X})$ ONC\\
$M_{\sun}$ & log(\es) & ~ & log(\es) \\
\hline
$<0.16$ & 29.57 & 12    & 29.00\\
0.16--0.25 & 29.60 & 3  & 29.15\\
0.25--0.50 & 29.85 & 3  & 29.60\\
0.50--1.00 & 29.90 & 10 & 30.05\\
1.00--2.00 & 30.17 & 18 & 30.45\\
2.00--3.00 & 30.30 & 10 & 31.20\\
3.00--10.0 & 30.42 & 1  & 30.50\\
$>10.0$ & 30.51 & 4     & 30.90\\
    \end{tabular}
    \label{tab:lx}
  \end{center}
\end{table}

\subsection{Spectral and timing analysis  of the brightest X-ray sources}

A spectral and timing analysis was carried out for the seven brightest
X-ray sources in the list of Table~\ref{tab:auto} (sources 22, 30, 32,
39, 60, 68, 72). Their light curves are shown in
Fig.~\ref{fig:lcspectra}. Sources 22 and 68 present large-amplitude
variability, and source 22 is not visible (no photons are detected)
for the first 10 ks of the \chandra\ exposure; the rise in the source
luminosity at 10 ks is impulsive and it is followed by a decay lasting
about 10 ks, with a behavior typical of stellar flares.  The mass of
source 22 is estimated at $M \simeq 0.6\,M_{\sun}$.  The time
variability of source 68 is remarkable, with the source counts
suddenly increasing to more than 10 times the previous level in about
2 ks and then decreasing abruptly again to a level about 3 times
higher than before the jump.  We do not provide an estimate for the
mass as its position in the color-magnitude diagram is inconsistent
with a 2 Myr isochrone at 600 pc. Both in the X-ray image and in the
2MASS data source 68 has an apparent companion, source 67, at about
$4.7$ arcsec. Both sources are visible in the X-ray soft band and
their $A_K$ derived with the approach described by LDK02 are very
similar, consistent with the two sources being physically associated.

The Kolmogorov-Smirnov (K-S) test, which measures the maximum
deviation of the integral photon arrival times from a constant source
model, was applied to the light curves, with the results summarised in
Table~\ref{tab:spectra}. As expected sources 22 and 68 have a
negligible probability of being constant. The probability of constancy
is also low for sources 30 and 39.

The background subtracted spectra for the seven brightest sources are
plotted in Fig.~\ref{fig:lcspectra}, together with the best-fit
models. The ACIS-I spectra, rebinned to a minimum of 5 counts per
energy bin, were fit with an absorbed one-temperature plasma model
with metal abundance 0.2 of solar, a value typical for T Tauri stars
(e.g. \citealp{fgm+03}). The best fit spectral parameters are
summarised in Table~\ref{tab:spectra}. The average coronal temperature
for all sources but source 68 (likely a foreground source) is
$kT=3.1$~keV, excluding also source 30 because of the large
uncertainty the average is $kT=2.4$~keV. From the spectral fits
intrinsic X-ray luminosities have been derived (also shown in
Table~\ref{tab:spectra}), assuming a source distance $d=600~$pc.

Excluding source 68, the values of absorbing column densities
derived for these sources are all higher than $5.0 \times 10^{21}$
cm$^{-2}$ (corresponding to $A_K>0.3$), consistent with the sources
being in a region of high extinction. Sources 30 and 32 have values of
\nh\ around the lower limit of the typical absorbing column density
for our sample of 61 sources, so that they could also be active
foreground stars. Nevertheless their $A_K$ is consistent with cluster
membership, as is their X-ray luminosity.

The best-fit values of \nh\ and $L_{\rm X}$ reported in
Table~\ref{tab:spectra} are consistent, within a factor of three for
the X-ray luminosity and within $2 \sigma$ for \nh, with the ones
computed using the photometrically derived \nh\ and assuming a typical
spectrum. This provides a consistency check on the method used to
derive \nh\ and $L_{\rm X}$ for all other sources in our sample of 61
for which a spectral analysis was not possible.

\begin{figure*}[!tbp]
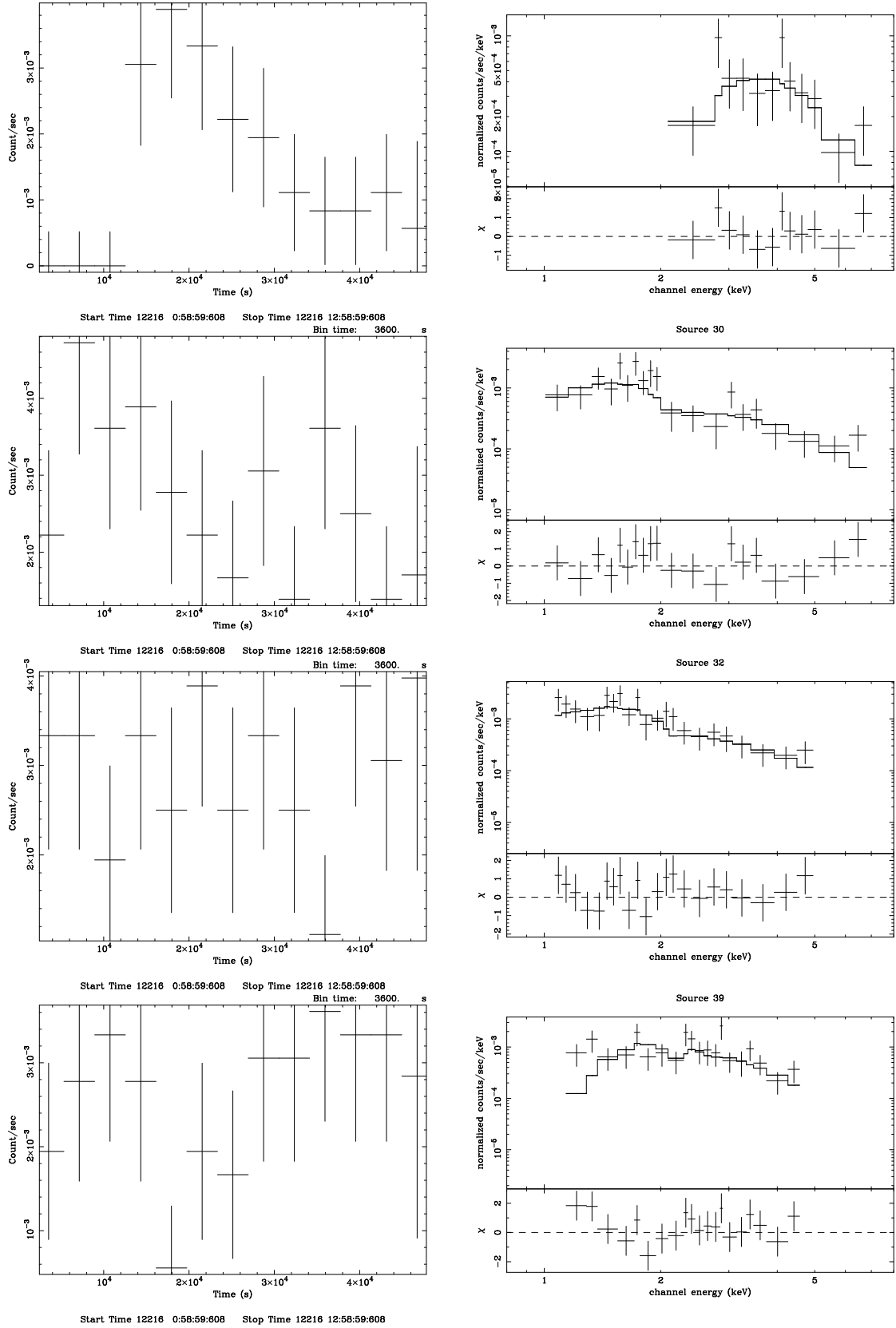

  \begin{center}\leavevmode
        \epsfig{file=0363fg06.ps, height=7.0cm, angle=270}\hspace{20pt}
        \epsfig{file=0363fg07.ps, height=7.0cm, angle=270}
        \epsfig{file=0363fg08.ps, height=7.0cm, angle=270}\hspace{20pt}
        \epsfig{file=0363fg09.ps, height=7.0cm, angle=270}
        \epsfig{file=0363fg10.ps, height=7.0cm, angle=270}\hspace{20pt}
        \epsfig{file=0363fg11.ps, height=7.0cm, angle=270}
        \epsfig{file=0363fg12.ps, height=7.0cm, angle=270}\hspace{20pt}
        \epsfig{file=0363fg13.ps, height=7.0cm, angle=270}

        \caption{ACIS-I light curves (left) and spectra (right) for
          the seven X-ray sources detected within a 4.4 arcmin radius
          from the center of S106, which are bright enough (count rate
          greater than 2 counts/ks) for a timing and spectral analysis
          to be carried out. The fits to the spectra with absorbed
          one-temperature plasma model are also shown.}
    \label{fig:lcspectra} \end{center}
\end{figure*}

\addtocounter{figure}{-1}
\begin{figure*}[!tbp]
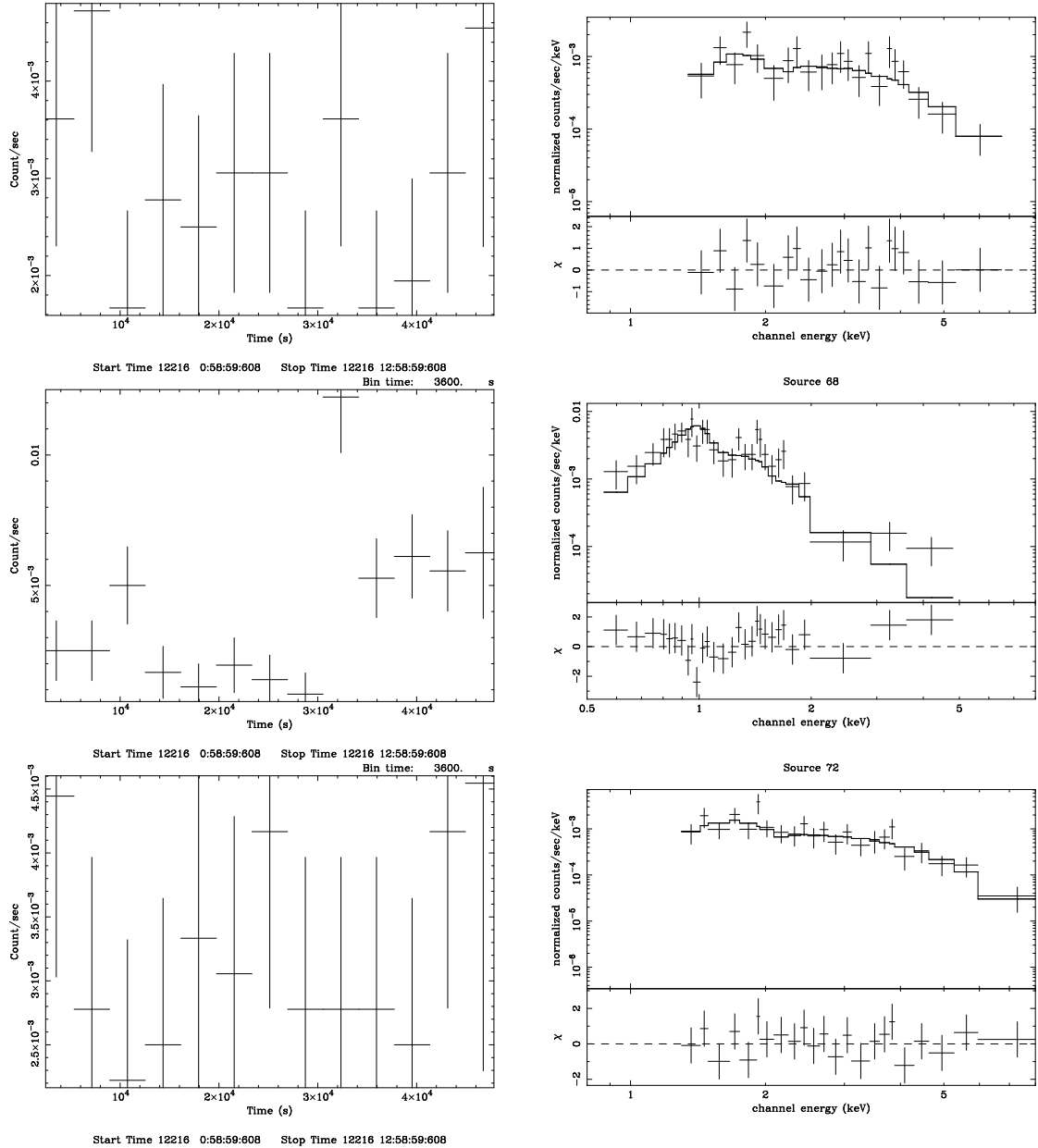

  \begin{center} 
    \leavevmode 
        \epsfig{file=0363fg14.ps, height=7.0cm, angle=270}\hspace{20pt}
        \epsfig{file=0363fg15.ps, height=7.0cm, angle=270}
        \epsfig{file=0363fg16.ps, height=7.0cm, angle=270}\hspace{20pt}
        \epsfig{file=0363fg17.ps, height=7.0cm, angle=270}
        \epsfig{file=0363fg18.ps, height=7.0cm, angle=270}\hspace{20pt}
        \epsfig{file=0363fg19.ps, height=7.0cm, angle=270}
\caption{\emph{continued --} light curves (left) and spectra (right) for
  the seven brightest X-ray sources in S106.}
  \end{center}
\end{figure*}

\begin{table*}[thbp]
  
   \caption{Summary of the derived X-ray properties for the 7
   brightest sources in our sample. \nh~phot. gives the absorbing
   column density as derived from the IR photometric data (see
   Sect.~\ref{sec:xrl} for the procedure). \nh and $kT$ give the best
   fit values for the absorbing column density and the plasma
   temperature as derived from spectral fitting of the ACIS-I data
   with one-temperature plasma models. The models' reduced $\chi^2$
   and null hypothesis probability are given in the two next
   columns. $F_{\rm x}$ and ``$F_{\rm x}$~unabs.'' give the observed
   and unabsorbed X-ray flux for the best fit model. $L_{\rm X}^1$ and
   $L_{\rm X}^2$ are, respectively, the luminosity derived from
   ``$F_{\rm x}$~unabs'' and the luminosity derived with the same
   approach as for all the other sources in our sample for which a
   spectral analysis was not possible. $M$ gives the stars' estimated
   mass, and K-S indicates the probability of constancy according to
   the Kolmogorov-Smirnov statistical test.  Unless otherwise
   specified, fluxes are in the energy range 0.8--7.5~keV. \nh\ is in
   units of $N_{22} = 10^{22}$~cm$^{-2}$, fluxes are in unit of
   $F_{-13}=10^{-13}~$\ecms\ and $L_{\rm X}$ in unit of $L_{30} =
   10^{30}$\es. }

\begin{center} 
\leavevmode
    \begin{tabular}{lccccccccccc}

\hline
\hline
Source  & $N({\rm H})$  phot. & $N({\rm H})$ &$kT$ & $\chi^2$ & $P$ &
$F_{\rm x}$ & $F_{\rm x}$ unabs. & $M$ & $L_{\rm X}^1$ & $L_{\rm X}^2$ & K-S\\
~       & $N_{22}$ & $N_{22}$ &keV & ~ & ~ & $F_{-13}$ & $F_{-13}$ & $M_{\sun}$ & $L_{30}$ & $L_{30}$\\
\hline
22 & 3.3 & 9.7$\pm$5.5 & 2.64$\pm$2.46 & 0.62 & 0.79 & 0.7$^a$ &
2.6$^a$ & 0.6 & 11 & 13 & 0.25$\times 10^{-7}$\\ 
30 & 1.3 & 0.8$\pm$0.4 & 6.71$\pm$5.57 & 1.00 & 0.46 & 0.3 & 0.3 & 1.4
& 1.3 & 3.1 & 0.20$\times 10^{-2}$\\
32 & 0.6 & 0.9$\pm$0.4 & 2.61$\pm$1.02 & 0.66 & 0.87 & 0.3 & 0.4 & 2.4
& 1.7 & 1.6 & 0.22\\
39 & 3.1 & 4.1$\pm$1.0 & 1.23$\pm$0.37 & 1.20 & 0.25 & 0.3 & 2.0 & 2.7
& 8.6 & 7.6 & 0.56$\times 10^{-2}$\\
60 & 3.3 & 3.2$\pm$0.8 & 2.33$\pm$0.75 & 0.72 & 0.82 & 0.4 & 1.0 & 2.9
& 4.3 & 9.2 & 0.83$\times 10^{-1}$\\
68$^b$ & - & 0.2$\pm$0.1 & 1.02$\pm$0.11 & 1.14 & 0.19 & 0.2 & 0.3 &
- & 0.4 & - & 0.59$\times 10^{-7}$\\
72 & 2.4 & 2.1$\pm$0.6 & 2.99$\pm$1.27 & 0.79 & 0.73 & 0.4 & 0.8 & 1.5
& 3.5 & 6.7 & 0.18\\

\hline \end{tabular} \label{tab:spectra} \end{center} $^a$ range
    1.4--7.5 keV\\ $^b$ likely a foreground source: it is inconsistent
    with a 2 Myr isochrone at 600 pc
\end{table*}

\section{Discussion}
\label{sec:disc}

\subsection{S106 membership}

In the \chandra\ image of the S106 star-forming region 87 X-ray
sources were detected within a 4.4 arcmin radius of the center of the
S106 cluster and 6 sources were detected within 0.8 arcmin radius of
the S106 south cluster. Of these, 71 sources appear to have an IR
counterpart in the 2MASS PSC catalogue and 22 have no IR counterpart
within 3 arcsec from the X-ray position.

The 22 sources without IR counterpart are likely stars, members of the
S106 cluster.  At a flux level of the order of 10$^{-14}$~\ecms\ the
expected number density of background sources determined on the basis
of the $\log N-\log S$ relationship for X-ray sources (see e.g.
\citealp{haa+01}) is 100--200 sources per square degree. The area of a
circle with 4.4 arcmin radius is 0.017 square degree, so that the
expected number of serendipitous extra-galactic sources for a low
absorption field is between 2 and 3. In practice given the high
absorption value in this field it is unlikely that any of these 22
sources is a background source (of galactic or extra-galactic origin).
The fact that none of these 22 sources shows up in the soft X-ray band
is an other indication that most of them are affected by significant
absorbing column density and therefore are unlikely to be foreground
sources.

From the sample of 71 X-ray sources with IR counterpart we selected a
subsample of 61 sources which have (photometrically derived) high
extinction levels ($\nh$ within a range of $[0.5-6.0] \times
10^{22}$~cm$^{-2}$) consistent with the sources being members of the
clusters. Of all these 61 sources only two (source 30 and source S106
south number 4) are visible in the X-ray soft band. This is another,
independent, indication of high column densities in front of most the
sources in the final sample. Of the two sources that were detected in
the soft band we were able to study source 30 in more detail, finding
that it has the characteristics of a pre-main sequence star (PMS) in
the outer edge of the association.  Thus, the 61 members of the final
sample are all likely members of the S106 star forming region.

\subsection{Age of the association}

For the final sample of 61 sources we studied the X-ray luminosity as
a function of mass. As apparent from Fig.~\ref{fig:LxM} and Table
\ref{tab:lx} there is a clear correlation between X-ray luminosities
and star masses. The presence of a correlation between X-ray
luminosity and star masses has been established for other star forming
regions (e.g.  \citealp{pz02} for IC~348, \citealp{fdm+03} for the
ONC) and it is a consequence of the fact that members of star-forming
regions are young and emit at similar, close-to-saturated levels of
X-ray luminosity. The presence of such a correlation in our sample
therefore indicates that the stars are young and physically
associated. A sample of field stars at random distances would not
present such a correlation.

For stars in the mass bins 0.5--2.0 $M_{\sun}$ we derive a median
value of $L_{\rm X}=8\times 10^{29}~$\es. Considering the age
evolution of the X-ray luminosity of solar-mass stars, and using the
values tabulated by \cite{micela02} for the median $L_{\rm X}$ of a
number of star forming regions, open clusters and nearby field stars,
one finds that a median value of $L_{\rm X}=8\times 10^{29}~$\es\ is
very similar to the median luminosities in the ONC, the $\rho$ Oph
star-forming region and the $\alpha$ Per cluster. This indicates for
the S106 clusters an age lower than $3\times 10^7$~yr.  In addition,
the median X-ray luminosity values derived here for the S106
star-forming region are, for all mass bins apart one ($2.0-3.0~M_{\sun}$),
within a factor of 3 of the values derived by \cite{fdm+03} for the
ONC, consistent with the age of S106 being similar to the one of the
ONC, as well as with the age estimate of \cite{hr91} (1--2$\times
10^{6}$ yr).

We note that the median value $L_{\rm X} \simeq 10^{30}$~\es\ for stars
in our sample with masses of 0.5--2.0\,$M_{\sun}$ is not strongly
dependent on the use of a 2 Myr year theoretical isochrone to estimate
extinction coefficients and masses. We have repeated the same
procedure using a zero age main sequence\footnote{Although this is an
incorrect assumption for an associations in which stars are nearly
coeval it allows the dependency of our results on the age assumption
to be tested.}  by \cite{sdf00} and the median luminosity derived for
stars within this mass range is essentially unchanged.  This is
because for stars with masses of 0.5--2.0\,$M_{\sun}$ in the 2 Myr hypothesis the
difference in mass estimate using a zero age main sequence is only a
factor $\sim 2$. The value of $L{\rm x}$ is affected by the age
assumption via the extinction coefficient and this estimate is also
not so sensitive to the use of a 2 million year isochrone or a zero
age main sequence. In the first case we derive a median value of $A_K
= 1.1$ and in the second case $A_K = 1.5$.

We caution however that in general the use of a zero age main sequence
rather than the appropriate isochrone can lead to significant
over-estimate of the cluster total mass. For instance in our case the
use of a zero age main sequence leads to 22 stars in our sample having
more than $3\,M_{\sun}$ while using a 2 Myr isochrone one only finds 5
stars. This may explain the significant difference in cluster mass
estimates by \cite{hr91} who derive a total mass of $140\,M_{\sun}$
and LDK02 who derive 400--600\,$M_{\sun}$.  LDK02 used a linear fit to
a zero age main sequence while \cite{hr91} adopted isochrones between
0.1--2.0 Myr.

\subsection{S106~IRS~4}

S106~IRS~4 is a massive stellar object of spectral type O7--B0 and
luminosity of [0.4--$1] \times 10^5~L_{\sun}$ (\citealp{ggc82};
\citealp{fms+84}). The line-of-sight extinction is estimated at
$A_V=20$--$30$ (\citealp{eel79}; \citealp{sfm+83}). From 2MASS
photometric data we estimated for S106~IRS~4 an extinction coefficient
$A_K = 3.35$, corresponding to $A_V = 30$. The infrared images are
consistent with a model in which the recently formed massive stars
excites the biconical nebula from within a large irregular disk of gas
and dust which is the remnant of the stellar collapse process
(\citealp{ggc82}, \citealp{hlj87}).  The near infrared flux from
S106~IRS~4 is probably not photospheric but results from scattered and
thermal radiation from the inner region of a circumstellar shell.

Radio observations show the existence of a compact shell of radio
emission surrounding S106~IRS~4 (\citealp{sfm+83}; \citealp{kcw94}).
This is interpreted as emission from a flowing ionized envelope from a
short-lifetime-phase of the pre-main sequence evolution of a massive
star (\citealp{sfm+83}; \citealp{fms+84}) -- an evolutionary state
with high mass loss which follows an earlier accretion phase.


The rate at which S106~IRS~4 is losing mass is atypically high for a
star of this bolometric luminosity. At $\dot{M}\sim 1.6 \times
10^{-6}\,M_{\sun}$\,yr$^{-1}$, corresponding to $\dot{M}/L \simeq
[1.6$--$8] \times 10^{-11} M_\odot\, L_\odot^{-1}$~yr$^{-1}$
(\citealp{fms+84}), this is 1--2 orders of magnitude higher than most
normal early type stars and comparable to the values measured in
Wolf-Rayet stars. At the same time the terminal wind velocity of
$v_{\infty} \sim 200$~km~s$^{-1}$ is lower than typical values
measured for equally luminous stars, for which $v_{\infty} \sim
1000$--$1500$~km~s$^{-1}$ (\citealp{pm82}). Radio observations
suggests that this wind may be mainly equatorial (\citealp{hdm+94}).
\cite{ssk+02} attribute the dynamics of the molecular gas in the S106
region to the impact of the ionized wind of S106~IRS~4, driving a shock
into an inhomogeneous molecular cloud.

The \chandra\ observation provides the first detection of S106~IRS~4
in X-rays. We estimated for this source an X-ray luminosity of $2
\times 10^{30}$\es\ assuming a plasma temperature $kT = 2.16$~keV, as
for all the other sources in our sample. This luminosity corresponds
to $L_{\rm X}/L_{\rm bol} \sim [0.5-1] \times 10^{-8}$, which is about
one order of magnitude lower than typically observed for older massive
stars (\citealp{svh+90}), although there is a large scatter ($\pm
1$~dex) on the $L_{\rm X}/L_{\rm bol}$ relation for massive stars
(e.g. \citealp{mcs+02}). The assumed X-ray plasma temperature for
S106~IRS~4 is however likely too high for an early type star. \cite{chs89}
have shown that typical plasma temperatures for wind related emission
in O stars are around $kT = 0.5$~keV, significantly lower than the
value we have assumed.  Given the high absorption toward the source,
the choice of plasma temperature is rather critical for determining
the X-ray luminosity. Assuming $kT = 0.5$~keV one derives for

S106~IRS~4 an intrinsic X-ray luminosity of $L_{\rm X} =4.8\times
10^{31}$ \es, corresponding to a range $L_{\rm X}/L_{\rm bol} \sim
[1.2$--$3] \times 10^{-7}$, at the lower end of the typical values for
O stars (\citealp{svh+90}). On the basis of the correlations
determined by \cite{svh+90} in their study of X-ray emission from
O-stars an X-ray luminosity of $4.8\times 10^{31}$ \es\ is a factor of
15 below the value typical for stars with comparable mass loss, but it
is similar to the X-ray luminosities found in relation to the star
wind momentum flux ($F_{\rm m} = \dot{M} v_{\infty} = 2\times
10^{27}~{\rm g~cm~s^{-2}}$).

\cite{kkh02} have recently reported the detection of X-ray emission
from four high-mass YSOs in Mon R2, deriving typical best fit plasma
temperatures, absorption column densities and X-ray luminosities of
$\sim 2$~keV, $\sim 5$--$10\times 10^{22}$~cm$^{-2}$ and
$10^{30}$--$10^{31}$\es, that is X-ray luminosities similar to the one
derived here for S106~IRS~4. The X-ray flux from the Mon R2 high mass
YSOs appears to be highly variable with flare-like behavior; because
of this and the high plasma temperatures \cite{kkh02} suggest that the
X-ray activity of these massive YSOs may be magnetically driven, in a
similar way to that seen in low mass PMS stars.

Unfortunately, for S106~IRS~4 no plasma temperature can be determined,
which means that we cannot compare it  with the values derived
for the Mon R2 sources and that we have to base the estimate of its
X-ray luminosity on an assumed plasma temperature.  Nevertheless,
assuming a plasma temperature typical for older massive stars, the
X-ray luminosity of S106~IRS~4 is consistent to the values predicted
for older stars on the basis of their wind momentum flux, which is the
dominating factor in determining the X-ray luminosity of a massive
star (as established by \citealp{svh+90} -- Fig. 16a). This suggests
that the activity in S106~IRS~4 maybe wind-driven, with no need to
invoke the presence of magnetically confined plasma. In S106~IRS~4, we
would thus be witnessing the onset of X-ray emission from the wind, at
a stage in which the protostar is still deeply embedded in the
circumstellar material. 

\subsection{Other interesting sources} 

Among the 7 sources for which we were able to study the X-ray spectra,
sources 32, 39 and 60 are intermediate mass stars, with estimated
masses of 2.4, 2.7 and $2.9\,M_{\sun}$, respectively. Were they main
sequence they would correspond to A stars, which are, at most,
weak X-ray sources ($L_{\rm X} \la 3\times 10^{27}$~\es, see e.g.
\citealp{fm03}). Given their young age however they could be Herbig
Ae/Be stars, which have significant X-ray activity, or even their
precursors, since according to theoretical models a two million year
old star of 2--3~$M_{\sun}$ will have spectral type K--G.

The luminosities ($L_{\rm X} = 2$--$9 \times 10^{30}$ \es) and plasma
temperatures ($kT \sim 2$~keV) of sources 32, 39 and 60 are
similar to those typically found for Herbig Ae/Be stars
(\citealp{pz96}; \citealp{hky+02}). While these values are not
inconsistent with the X-ray emission being from unseen low-mass
counterparts, evidence is gathering that the origin of the emission
are the Herbig Ae/Be stars themselves (\citealp{pz96};
\citealp{hky+02}; \citealp{gfm+04}).

Recently \cite{bkp+02} have reported X-ray emission from four
intermediate mass YSOs in the massive star forming region IRAS
19410+2336 which have comparable X-ray luminosity ($L_{\rm X} =
10^{31}$--$10^{32}$~\es) and high plasma temperature ($kT \ga 2$~keV).

\section{Conclusions}
\label{sec:concl}

The \chandra\ X-ray observation, combined with the 2MASS data, has
allowed the S106 and S106 south clusters to be studied in more detail
confirming them as sites of recent star formation, with age comparable
to that of the ONC. In addition, the X-ray observation has allowed
the low-mass YSO population of S106 to be identified, opening the way for
the IR study of the individual stars. The X-ray characteristics of
this population appear to be similar to the ones of the (much better
characterized) ONC.

We have detected X-ray emission from S106~IRS~4, a highly embedded
massive young stellar object with an exceptional wind for its
luminosity. While somewhat dependent on the plasma temperature (which
cannot be determined from our data), the X-ray luminosity of S106~IRS~4
appears low in comparison with massive main sequence and more evolved
massive stars, both on the basis of its bolometric luminosity as well
as its mass loss rate. Nevertheless, when wind momentum flux is
considered, which includes the influence of both the mass loss rate
and the terminal velocity, the X-ray luminosity of S106~IRS~4 appears
typical of more evolved massive stars, suggesting the same process to
be at work. The lower luminosity with respect to stars of similar mass
loss rate is due to the peculiarly low wind terminal velocity. This
would therefore suggest that the wind-driven X-ray emission of massive
stars is already present during this early pre-main sequence phase.
As discussed e.g. by \cite{bkr+98}, S106~IRS~4 appears to be extremely
young, and \cite{fms+84} argue that the star is in a short-lived phase
of its early evolution which follows an earlier accretion phase and it
is characterized by high mass loss.

Further observations of S106~IRS~4 and other objects of this type (the
present observation being thus far unique) will be required to
establish whether such an early start of the X-ray emission in massive
stars is a common phenomenon and whether the emission is wind-driven
already at these early stages. Were this the case one could speculate
that a wind-driven X-ray emission in S106~IRS~4 is a consequence of it
being in a more advanced evolutionary phase in respect to the massive
YSOs detected in the X-ray by \cite{kkh02} and for which they argue
that a magnetic confined plasma is present.

\begin{acknowledgements}
  
  GM acknowledges financial support from ASI.  This research has made
  use of the NASA/ IPAC Infrared Science Archive (operated by the Jet
  Propulsion Laboratory, California Institute of Technology, under
  contract with the National Aeronautics and Space Administration) and
  of the \chandra\ archive.
 
\end{acknowledgements}

\bibliographystyle{aa}

\end{document}